# Development of a computer-aided diagnostic system for Alzheimer's disease using magnetic resonance imaging

Kenya Murase[*], Naohiko Gondo, and Tsutomu Soma

*Department of Medical Physics and Engineering, Division of Medical Technology and Science, Faculty of Health Science, Graduate School of Medicine, Osaka University*

*1-7 Yamadaoka, Suita, Osaka 565-0871, Japan*

[*]Address correspondence to:

    Kenya Murase, Dr. Med. Sci., Dr. Eng.

    Department of Medical Physics and Engineering, Division of Medical Technology and Science, Faculty of Health Science, Graduate School of Medicine, Osaka University

    1-7 Yamadaoka, Suita, Osaka 565-0871, Japan

    Tel & Fax: (81)-6-6879-2571,

    e-mail: murase@sahs.med.osaka-u.ac.jp

# ABSTRACT

Alzheimer's disease (AD) is the most common type of dementia accompanied with brain atrophy. Structural measurements of brain atrophy in specific brain structures such as hippocampus using magnetic resonance imaging (MRI) have been reported to detect the development of dementia early in the course of the disease. The purpose of this study was to develop a computer-aided diagnostic system for AD using MRI, which is based on the automatic volumetry of segmented brain images and generation of three-dimensional cortical thickness images using the Eulerian partial differential equation (PDE) approach. We investigated the effect of the inhomogeneity of magnetic field strength and statistical noise on the accuracy of our automatic volumetry and the PDE approach using the simulated MR images generated from BrainWeb. Our automatic volumetry and PDE approach were robust against inhomogeneous magnetic field strength. Although the accuracy of our automatic volumetry decreased with increasing statistical noise, it was maintained when the statistical noise was less than 7-8%. The cortical thickness obtained by the PDE method tended to decrease with increasing statistical noise. When we applied our method to clinical data, the cortical thinning due to brain atrophy was clearly demonstrated in patients with brain atrophy. These results suggest that our system appears to be useful for the diagnosis of AD, because it allows us to automatically evaluate the extent of brain atrophy and cortical thinning in a three-dimensional manner.

# I. INTRODUCTION

Alzheimer's disease (AD) is the most common type of dementia accompanied with brain atrophy [1]. Using magnetic resonance imaging (MRI), structural measurements of brain atrophy in specific brain structures such as hippocampus have been reported to detect the development of dementia early in the course of the disease [2]. The purpose of this study was to develop a computer-aided diagnostic system for AD using MRI, which is based on the automatic volumetry of segmented brain images and generation of three-dimensional (3D) cortical thickness images.

# II. MATERIALS AND METHODS

*Automatic volumetry of segmented brain images*

Figure 1 illustrates our process for obtaining subject's template. A subject's magnetic resonance (MR) image was first mached to anterior commissure – posterior commissure (AC-PC) plane of standard Montreal Neurological Institute (MNI) space (Simulated Brain Database) using the Statistical Parametric Mapping 99 (SPM99) software [3] with 6-parametric rigid body transformation. The transformation of each subject's image was performed by spatial normlization technique of SPM99 [3], using 12-parametric linear affine and further 12 non-linear iteration algorithms with $7 \times 8 \times 7$ basis functions and the segmented gray matter (GM) template smoothed with an isometric 8-mm full width at half maximum (FWHM) Gaussian kernel. Then, the aligned MR image was segmented into GM, white matter (WM) and cerebrospinal fluid (CSF) with SPM99 [3].

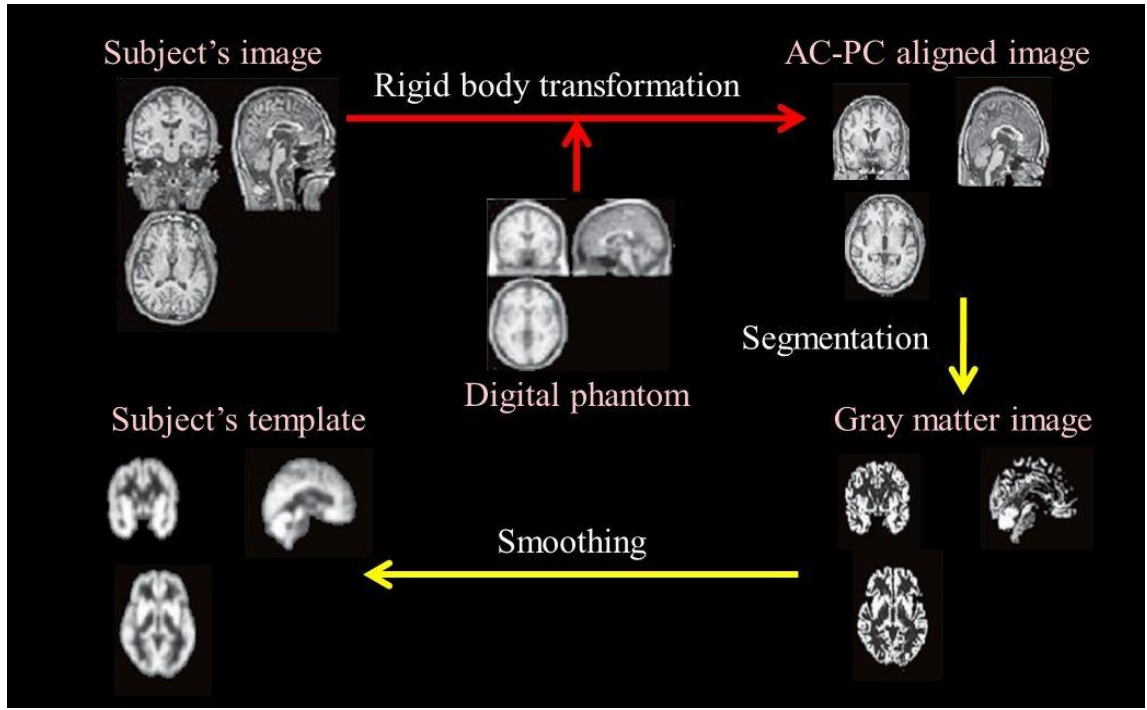

**Figure 1: Process for obtaining subject's template.**

The intarcranial area and hippocampus for the subject were detremined from the deformed volumes of interest (VOIs) which were defined on a stnadardized stereotactic space based on the MNI space and trasnformed to the subject's space (Fig. 2). The area of total intracranial volume (TIV) was adjusted with an image made from the segmented GM, WM and CSF images. The segmented images were calculated and 3D Gaussian convolution was performed with an isometric 6-mm FWHM, further binarized with the threshold set to 40% of the maximum value. A temporary GM area was obtained by an image which was binarized in the segmented GM image with the threshold set to 65% of the maximum value.

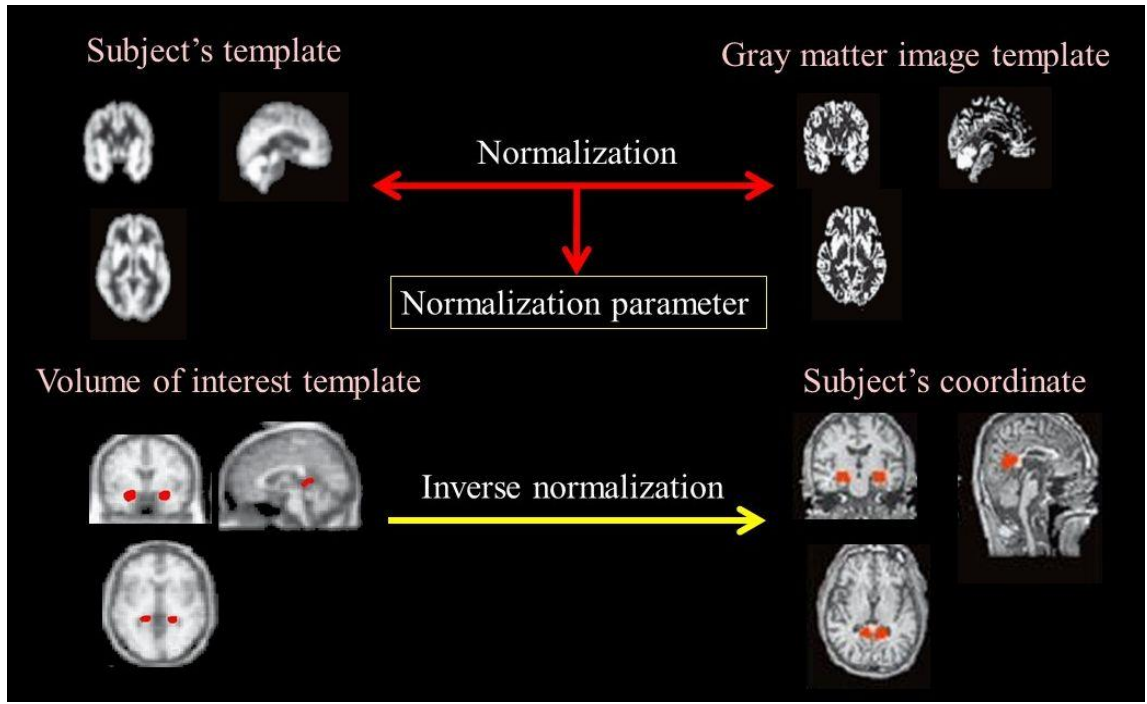

**Figure 2: Process for obtaining inverse anatomical standardization parameters.**

The WM area was also binalized from the segmented WM image with the threshold set to 35% of the maximum value, and filled defected areas. Next, the areas of GM and WM were calculated by the summation of the temporary GM and WM areas and filled defected arears. The GM area was reconstructed by subtraction from the temporary GM and WM area by the WM area. The TIV was calculated from the number of voxels in the area of the trasnformed intracranial VOIs. The volume of the GM and WM was calculated as the number of voxels in the obtained GM and WM area. The volume of the GM was calculated from the summation of voxel's value on the segmented GM image in the GM area. The volume of the WM was subtracted from the volume of the GM and WM. The volume of the GM in the hippocampus was calculated in the transformed hippocampus VOI template for the subject.

To investigate the accuracy of our automatic volumetry of segmented image method, we compared the TIVs, whole brain volumes (WBVs) and hippocampal volumes obtained by our method and those measured by the manual method.

*Effects of inhomogeneity of magnetic field strength and statistical noise*

To investigate the effects of the inhomogeneity of magnetic field strength and statistical noise on the accuracy of our automatic volumetry of segmented brain images, we generated the simulated images with inhomogeneity ranging from 0 to 100% and statistical noise ranging from 0 to 20% using BrainWeb (http://www.bic.mni.mcgill.ca/brainweb/). Figure 3 illustrates how to generate the simulated MR images with inhomogeneous magnetic field strength.

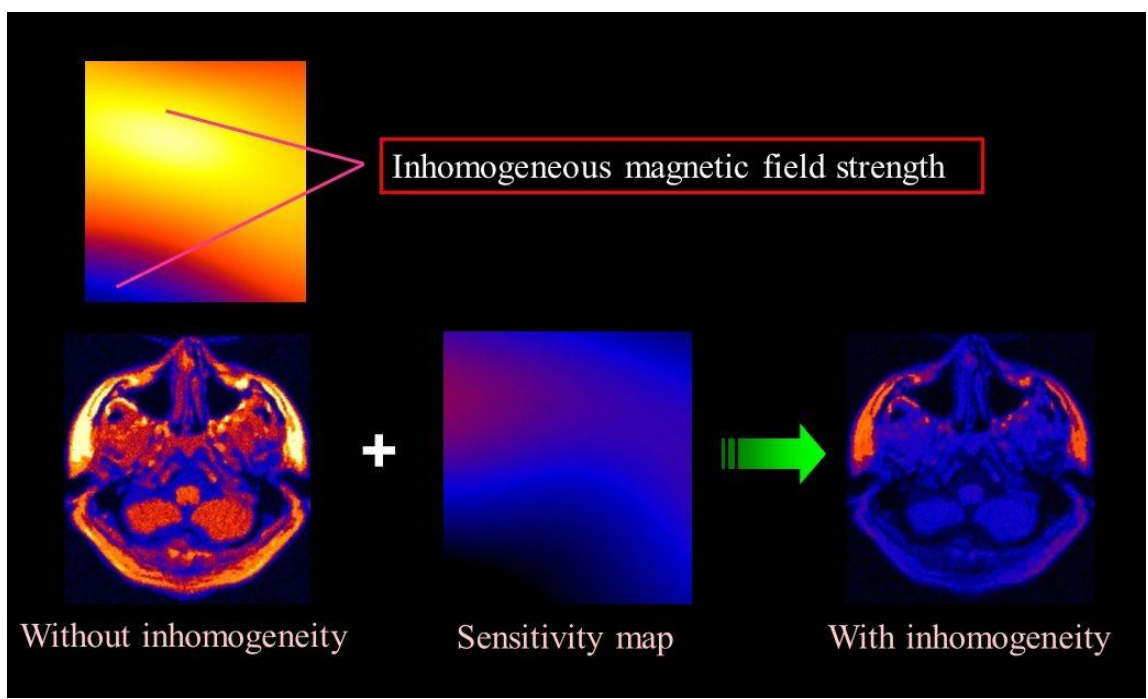

**Figure 3: Method for generating MR images with inhomogeneous magnetic field strength.**

*Generation of three-dimentional cortical thickness images*

The cortcial thickness was calculated using an Eulerian partial differential equation (PDE) approach [4] form the GW, WM and CSF images segmented using the method mentioned above. Then, the 3D image of the cortical thickness was generated. Figure 4 illustartes the Eulerian PDE approach for computing the cortical thickness [4].

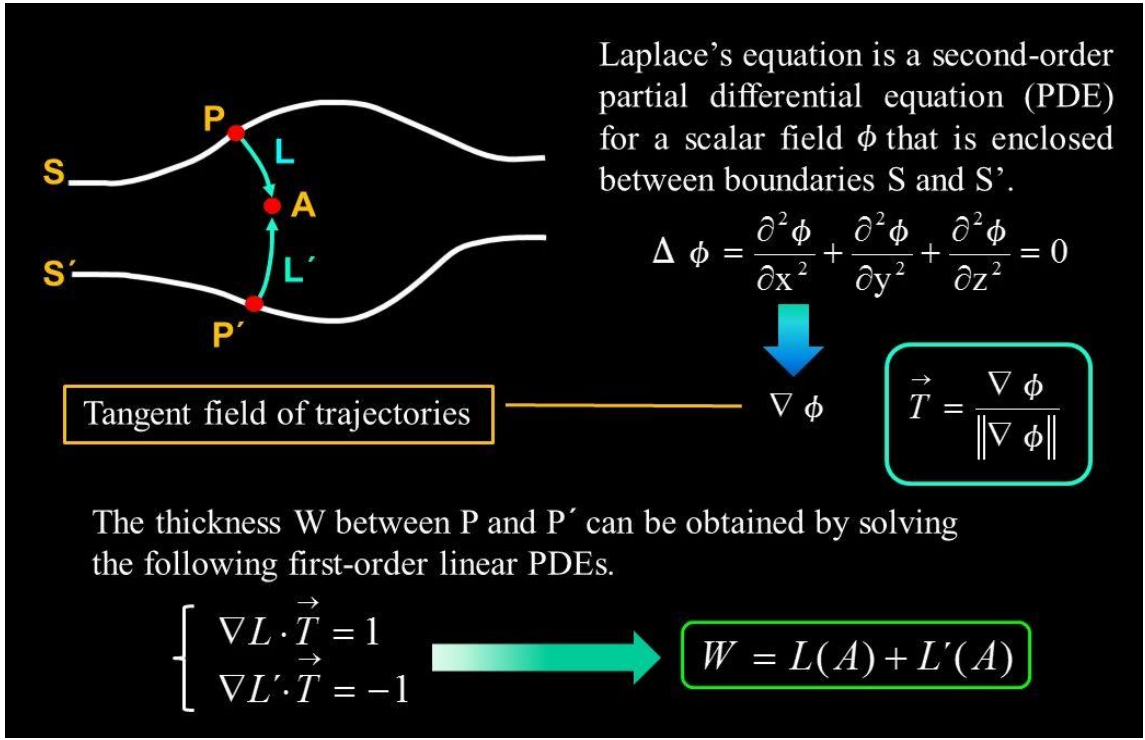

**Figure 4: Illustration of the Eulerian PDE approach for computing cortical thickness.**

### III. RESULTS

The TIVs, WBVs and hippocampal volumes obtained by our method correlated well with those measured by the manual method. The correlation coefficients were 0.910, 0.902 and 0.918 for TIVs, WBVs and hippocampal volumes, respectively (plot not shown).

Prior to the clinical application of the Eulerian PDE approach, we investigated its accuracy using numerical phantoms. Figure 5 shows an example of the numerical phantom and the thickness image obtained from the numerical phantom using the Eulerian PDE approach. Figure 5 also shows the correlation between the thickness of the numerical phantom actually measured (x, pixel) and that calculated using the Eulerian PDE approach (y, pixel) together with the Bland-Altman plot [5]. There was an excellent correlation between them (r=0.995, y=0.928x+2.168).

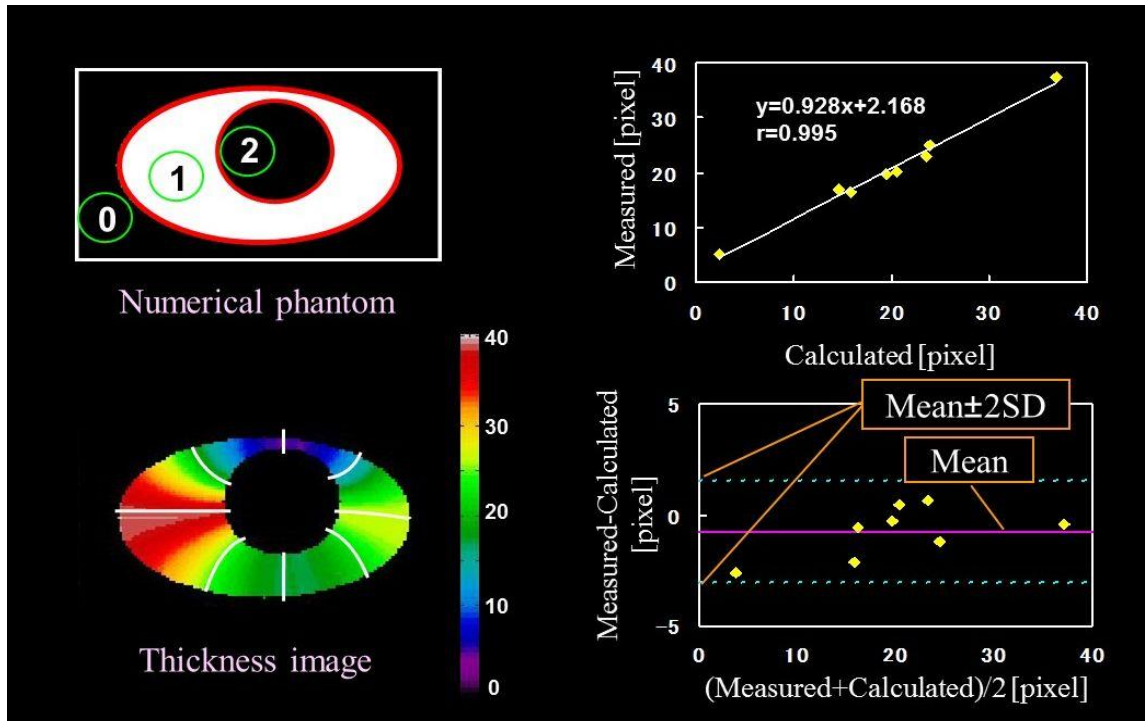

**Figure 5: Investigation on the accuracy of the Eulerian PDE approach using numerical phantoms.**

Figure 6 shows the results of the investigation of the effect of inhomogeneous magnetic field strength on the accuracy of our automatic volumetry of segmented brain images using the simulated MR images generated using BrainWeb (http://www.bic.mni.mcgill.ca/brainweb/).

These results suggest that our automatic volumetry is robust against inhomogeneous magnetic field strength.

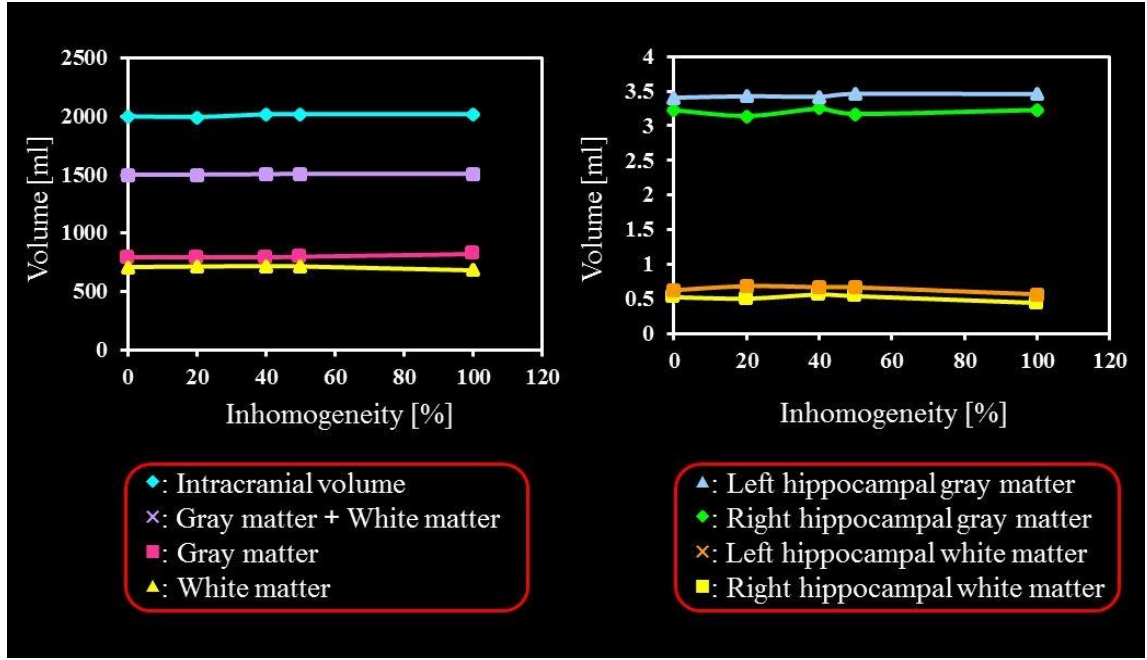

**Figure 6: Effect of inhomogeneous magnetic field strength on the accuracy of our automatic volumetry.**

Figure 7 shows the 3D display of the cortical thickness calculated from the simulated MR images using the Eulerian PDE approach for various values of inhomogeneous magnetic field strength. These results suggest that the Eulerian PDE approach is robust against inhomogeneous magnetic field strength.

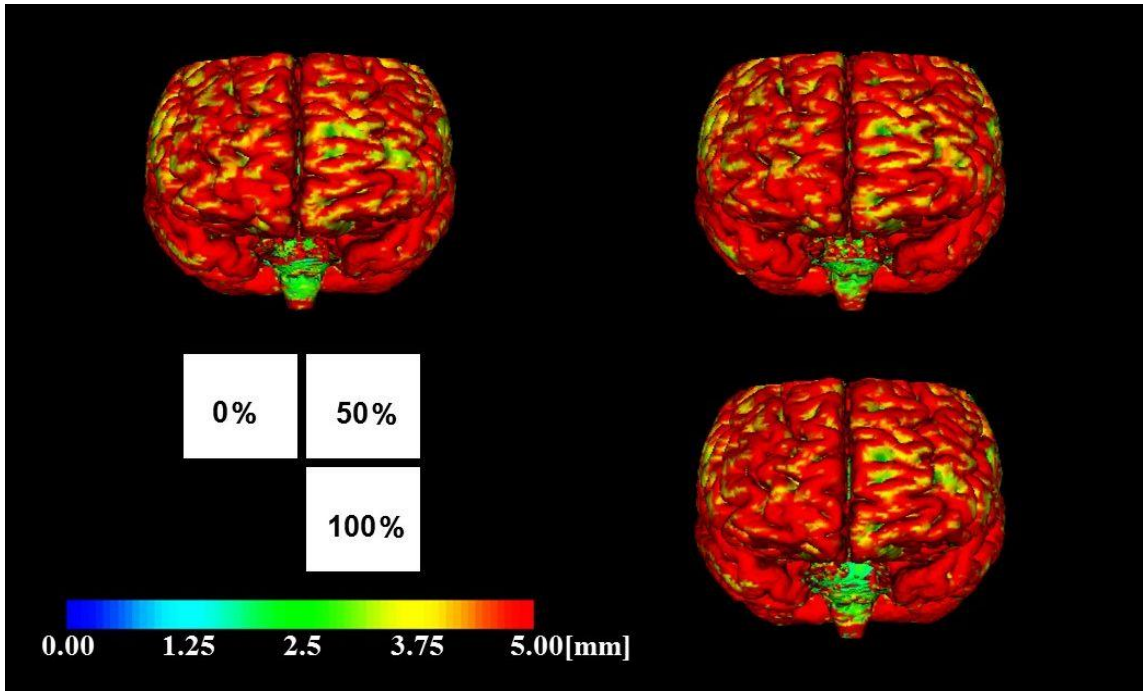

**Figure 7: Three-dimensional display of the cortical thickness calculated from the simulated MR images using the Eulerian PDE approach for various values of inhomogeneous magnetic field strength.**

Figure 8 shows the results of the investigation of the effect of statistical noise on the accuracy of our automatic volumetry of segmented brain images using the simulated MR images generated using BrainWeb (http://www.bic.mni.mcgill.ca/brainweb/). Although the accuracy of our method decreased with increasing statistical noise, the accuracy was maintained when the statistical noise was less than 7-8%.

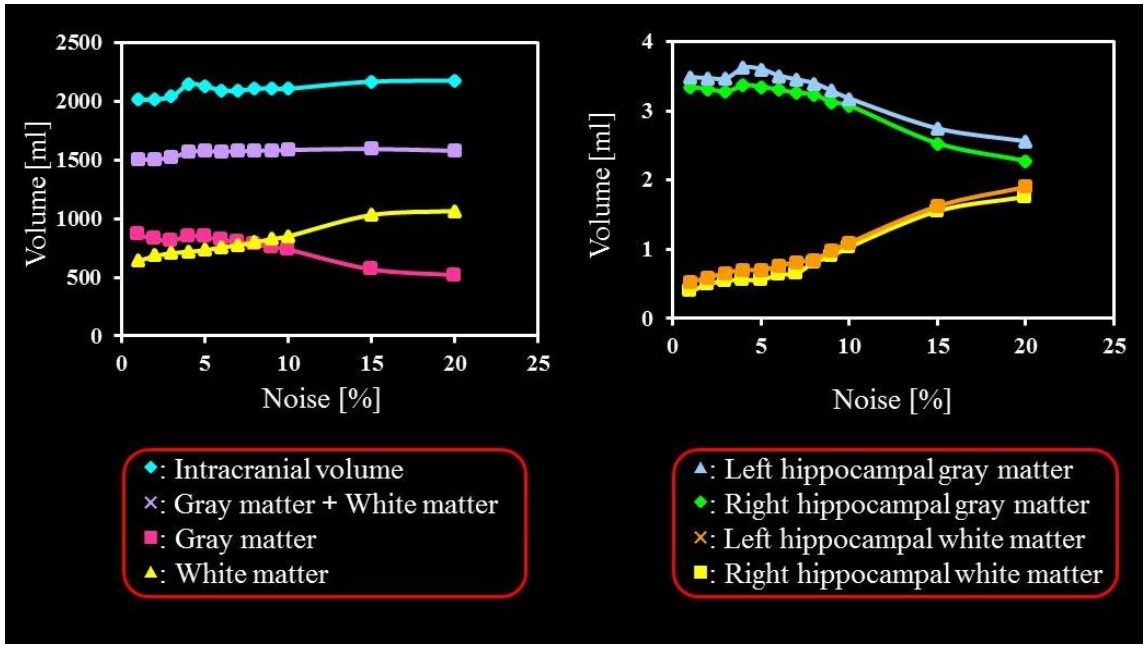

**Figure 8: Effect of statistical noise on the accuracy of our automatic volumetry.**

Figure 9 shows the 3D display of the cortical thickness calculated from the simulated MR images using the Eulerian PDE approach for various values of statistical noise. As shown in Fig. 9, there was a tendency for the cortical thickness to decrease with increasing statistical noise.

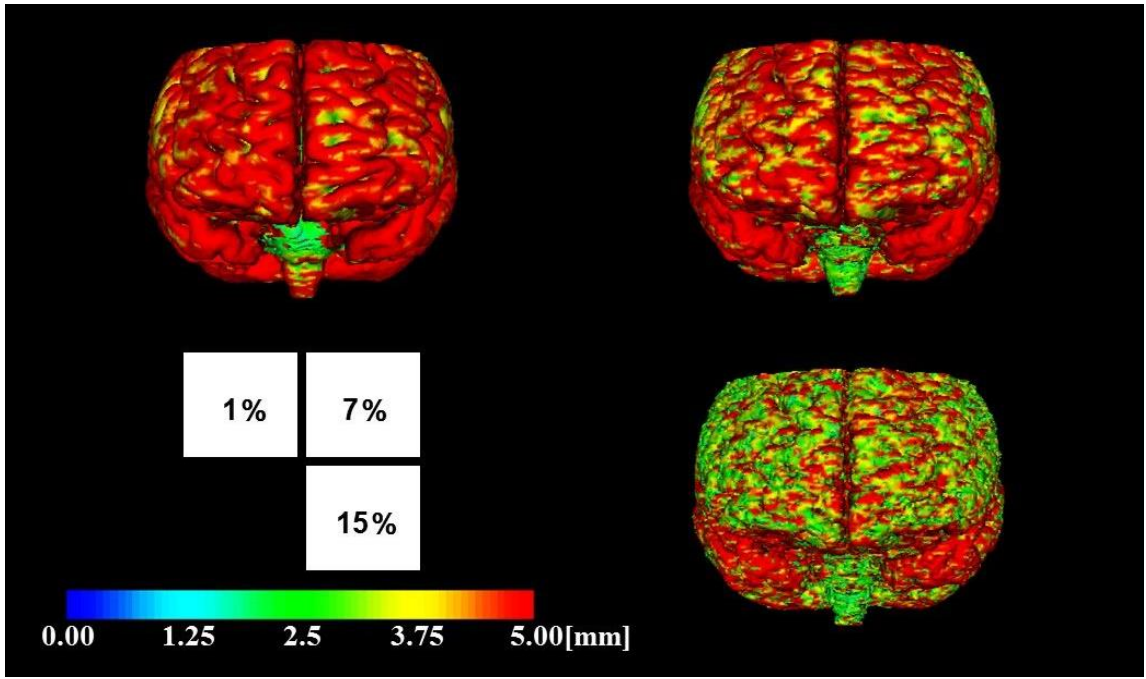

**Figure 9: Three-dimensional display of the cortical thickness calculated from the simulated MR images using the Eulerian PDE approach for various values of statistical noise.**

Figure 10 shows examples of the cortical thickness images obtained from patients without (left) and with brain atrophy (right) using the Eulerian PDE approach. As shown in Fig. 10, the cortical thinning due to brain atrophy was clearly demonstrated in patients with brain atrophy.

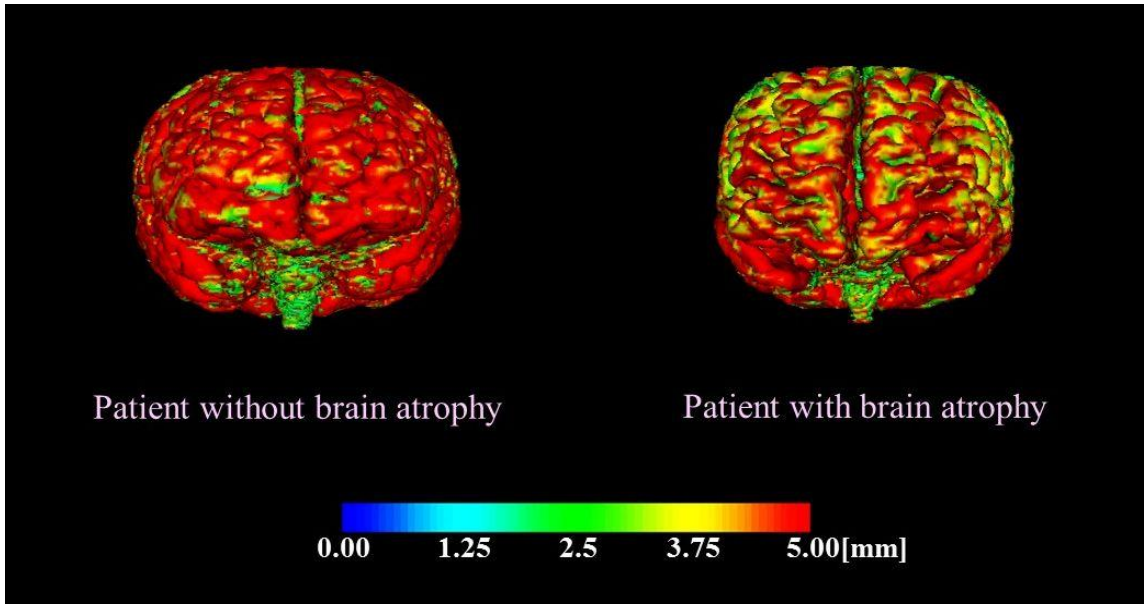

**Figure 10: Three-dimensional display of the cortical thickness calculated using the Eulerian PDE approach in patients without (left) and with brain atrophy (right).**

## IV. DISCUSSION

We have developed and validated a fully automatic algorithm for the measurement of TIV, WBV, hippocampal volume, and cortical thickness. Furthermore, we have developed a method for displaying cortical thickness in a three-dimensional manner. Our approach is based on combining a segmentation, anatomical normalization technique, VOI measurement, and Eulerian PDE approach.

For volumetric measurements, segmentation of GM, WM, and CSF is important because the actual value of the volume depends on how correctly each area is segmented [6,7]. A procedure for segmentation of these areas has been developed by several investigators. They improved brain segmentation by obtaining iterative tracking of the brain and calculating probability maps based on a model of intensity probability distribution that includes two partial volume classes [6,7], whereas SPM99's segmentation algorithm utilizes *a priori* probability images of GM, WM, and CSF. Before segmentation, each voxel in the image is mapped to its equivalent location in the *a priori* probability images through spatial transformation. The iterative segmentation algorithm is based on the maximum likelihood "mixture model" clustering algorithm [8]. The algorithm is terminated when the change in log-likelihood from the previous iteration becomes negligible [8].

As previously described, our fully automatic volumetry system adopted the SPM99 normalization and segmentation technique and we verified the accuracy of the anatomical normalization by the SPM99, especially hippocampal normalization. The limitation of our system is that it cannot be applied to severely atrophied brains because of a normalization limitation of the SPM99's program; however, in clinical use there is no need to measure brain volumes of severe AD patients. In this study, we focused on AD and measured the hippocampal volume; however, by producing another VOI template, we can measure other structures such as brain stem and cerebellum in patients with spinocerebellar degeneration or other degenerative diseases. In the future, we can expect to verify that this system can make further important contributions to the measurement of fine brain structures in patients with early AD, mild cognitive impairment, and other degenerative diseases.

## V. CONCLUSIONS

We developed a computer-aided diagnostic system for AD using MRI, which is based on automatic volumetry of segmented brain images and generation of three-dimensional cortical thickness images. Our system appears to be useful for the diagnosis of AD, because it allows us to automatically evaluate the extent of brain atrophy and cortical thinning in a three-dimensional manner.

## ACKNOWLEDGEMENTS

The authors are grateful to Prof. Teruhito Mochizuki, Dr. Keiichi Kikuchi, and Dr. Hitoshi Miki of Ehime University School of Medicine for allowing us to use clinical data. The authors are also grateful to Mr. Youichi Yamazaki for his help in developing the present system.